\documentclass[10pt,preprint]{aastex}

\usepackage{geometry}
\usepackage{graphicx}
\usepackage[symbol]{footmisc}
\usepackage{epstopdf}
\usepackage{subfigure}
\usepackage{amsmath}
\usepackage{float}
\usepackage{color}
\def\ms{\rm\,m\;s^{-1}}
\def\kms{\rm\,km\;s^{-1}}
\def\cms{\rm\,cm\;s^{-1}}

\def\half{\hbox{$\frac12$}}
\def\O{\mathcal O}
\def\H{H}
\def\star{{\textrm{star}}}
\def\photon{{\textrm{\thinspace null}}}

\def\Mbh{M_{\hbox{\tiny{BH}}}}

\def\static{{\textrm{static}}}
\def\kep{{\textrm{Kep}}}
\def\torq{{\textrm{q}}}
\def\schw{{\textrm{Schw}}}
\def\fd{{\textrm{FD}}}

\def\mink{{\textrm{Mink}}}

\begin{document}

\title{Towards relativistic orbit fitting of Galactic center stars and pulsars}

\author{Raymond Ang\'elil and Prasenjit Saha}

\affil{Institute for Theoretical Physics, University of Z\"urich, \\
Winterthurerstrasse 190, CH-8057 Z\"urich, Switzerland}

\and

\author{David Merritt}

\affil{Department of Physics and Center for Computational Relativity and
Gravitation, \\
Rochester Institute of Technology,\\
Rochester, NY 14623}

\begin{abstract}
  The S~stars orbiting the Galactic center black hole reach speeds of up to
  a few percent the speed of light during pericenter passage.  This makes,
  for example, S2 at pericenter much more relativistic than known
  binary pulsars, and opens up new possibilities for testing general
  relativity.  This paper develops a technique for fitting
  nearly-Keplerian orbits with perturbations from Schwarzschild curvature,
  frame dragging, and the black hole spin-induced quadrupole moment, to redshift measurements
  distributed along the orbit but concentrated around pericenter.
  Both orbital and light-path effects are taken into account.
  It turns out that absolute calibration of rest-frame frequency is
  not required.  Hence, if pulsars on orbits similar to the S~stars are
  discovered, the technique described here can be applied without change,
  allowing the much greater accuracies of pulsar timing to be taken
  advantage of. For
  example, pulse timing of $3\,\mu\rm s$ over one hour amounts to an
  effective redshift precision of $30\cms$, enough to measure frame
  dragging and the quadrupole moment from an S2-like orbit, provided
  problems like the Newtonian ``foreground'' due to other masses can be
  overcome.  On the other hand, if stars with orbital periods of
  order a month are discovered, the same could be accomplished with 
  stellar spectroscopy from the E-ELT at the level of $1\kms$.
\end{abstract}

\pagebreak

\section{Introduction}

Tracing the orbits of the S~stars in the Galactic center region 
reveals a central mass of $\sim 4 \times 10^6
M_\odot$ \citep{ghez,gillessen}, presumably a supermassive black 
hole.  The roughly 20 known S~stars are mostly main-sequence B stars, 
and it is likely that they share their environment with many fainter stars.  Hence discovery by the next generation of telescopes of stars far closer
to the black hole is anticipated.
The dynamics of the orbit and light
trajectories provide an opportunity to test hitherto unobserved
predictions of general relativity in the black hole neighborhood, and
in doing so, confirm or revise our understanding of the nature of
gravity  \citep{2010arXiv1002.1224G}. 
Meanwhile, searches for pulsars in the Galactic-center
region are also underway \citep[e.g.,][]{marquart}.  If pulsars on
orbits similar to S~stars are discovered, their time-of-arrival
measurements may provide accuracies orders of magnitude beyond
those available from optical spectroscopy.

\begin{figure}[h!]
\includegraphics[scale=0.93]{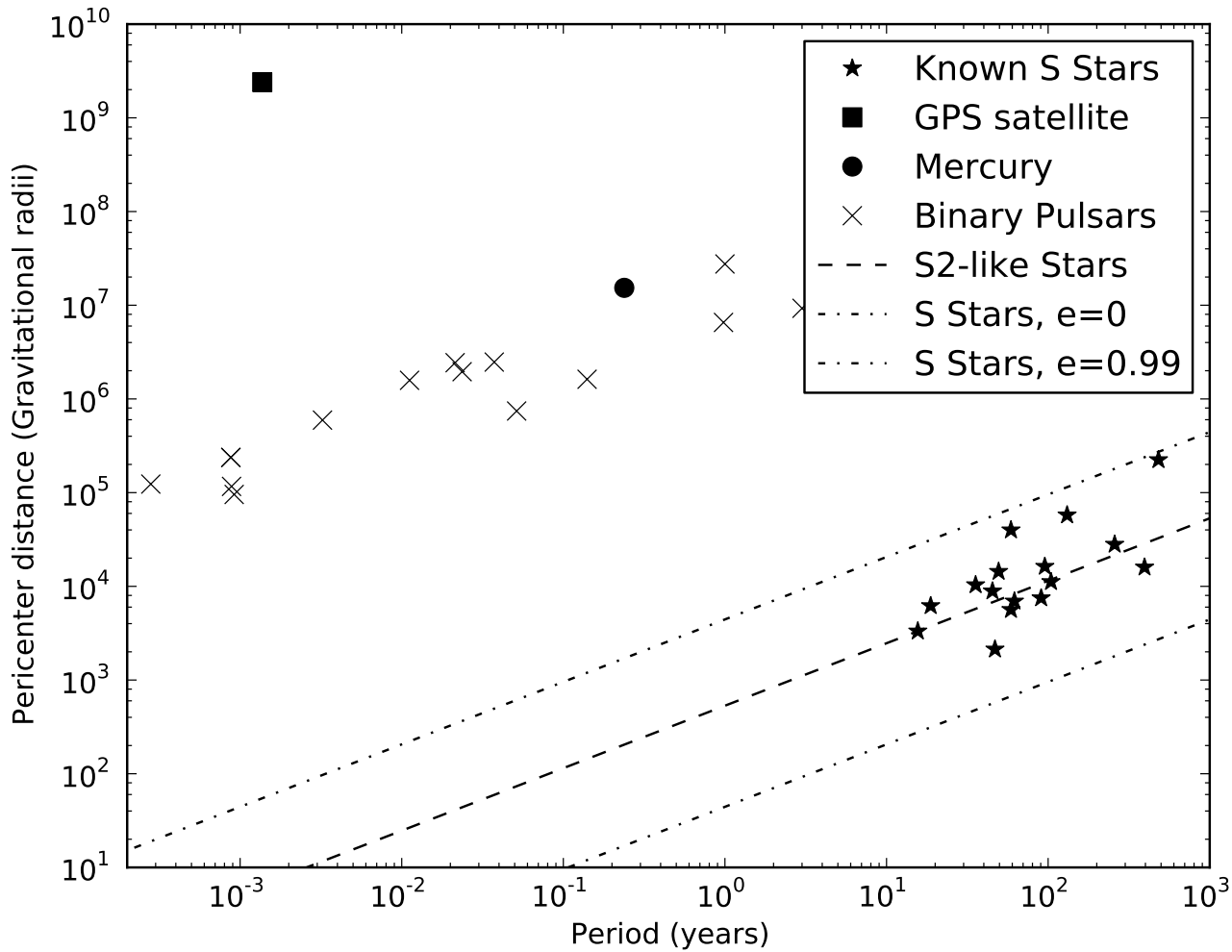}
\caption{Pericenter distance (in units of $GM/c^2$) against orbital
  period for a variety of systems. The known S~stars, having smaller
  pericenter distances, are more relativistic than known binary
  pulsars.  But the long orbital periods of the S~stars render it
  infeasible to measure cumulative effects over many orbits.  Hence
  other techniques must be devised.  The pulsar examples are taken
  from \citet{pulsar_table}. For the S~stars, the orbital elements in
  \citet{gillessen} have been used. In this paper we also treat 
  fictitious stars that lie along the dashed line, that is, having a
  range of semi-major axis $a$ values but with the same eccentricity and angular elements as S2.}
\label{fig:diff_beast}
\end{figure}

It is interesting to compare S~stars with other relativistic orbits. General relativistic perturbations of a nearly Keplerian orbit
typically scale as an inverse power of the classical angular momentum
$\sqrt{a(1-e^2)}$.  So a simple measure of the strength of relativistic
effects in a nearly Keplerian orbit is the
dimensionless pericenter distance.\footnote{In
  this paper the semi-major axis $a$ and the pericenter distance
  $a(1-e)$ are always expressed in units of $GM/c^2$ and are therefore
  dimensionless.}
Figure \ref{fig:diff_beast} plots pericenter distance against 
orbital period for a sample of S~stars and binary pulsars, together 
with Mercury and artificial satellites.  At present, binary pulsars are the
leading laboratories for general relativity.  The principles are
summarized in \cite{1994RvMP...66..711T}.  Taylor's Figure~11, depicting 
five intersecting curves, indicating four distinct tests of general
relativity, is especially memorable, and descendants of this
figure with newer data appear in recent work \citep[e.g., Figure 3
in][]{2009CQGra..26g3001K}.  So it is remarkable that the star S2 has
a pericenter distance $\sim 3\times10^3$, an order of magnitude
lower than the most relativistic binary-pulsar systems, and four
orders of magnitude lower than Mercury. This suggests that
Galactic center orbits may show relativistic effects not measured in
binary pulsars, such as frame dragging.  On the other hand, S2 has an
orbital period of $\sim 15.9\rm\,yr$ versus one day or less for some
binary pulsars.  Thus, for the known S~stars, or pulsars on similar
orbits, we cannot build up a relativistic signal over many orbits.  A
different strategy is needed, based on detailed observation of one or
a few orbits, and paying special attention to pericenter passage, when
relativistic effects are strongest.

\begin{center}
\begin{figure}[h!]
\includegraphics[scale=0.5]{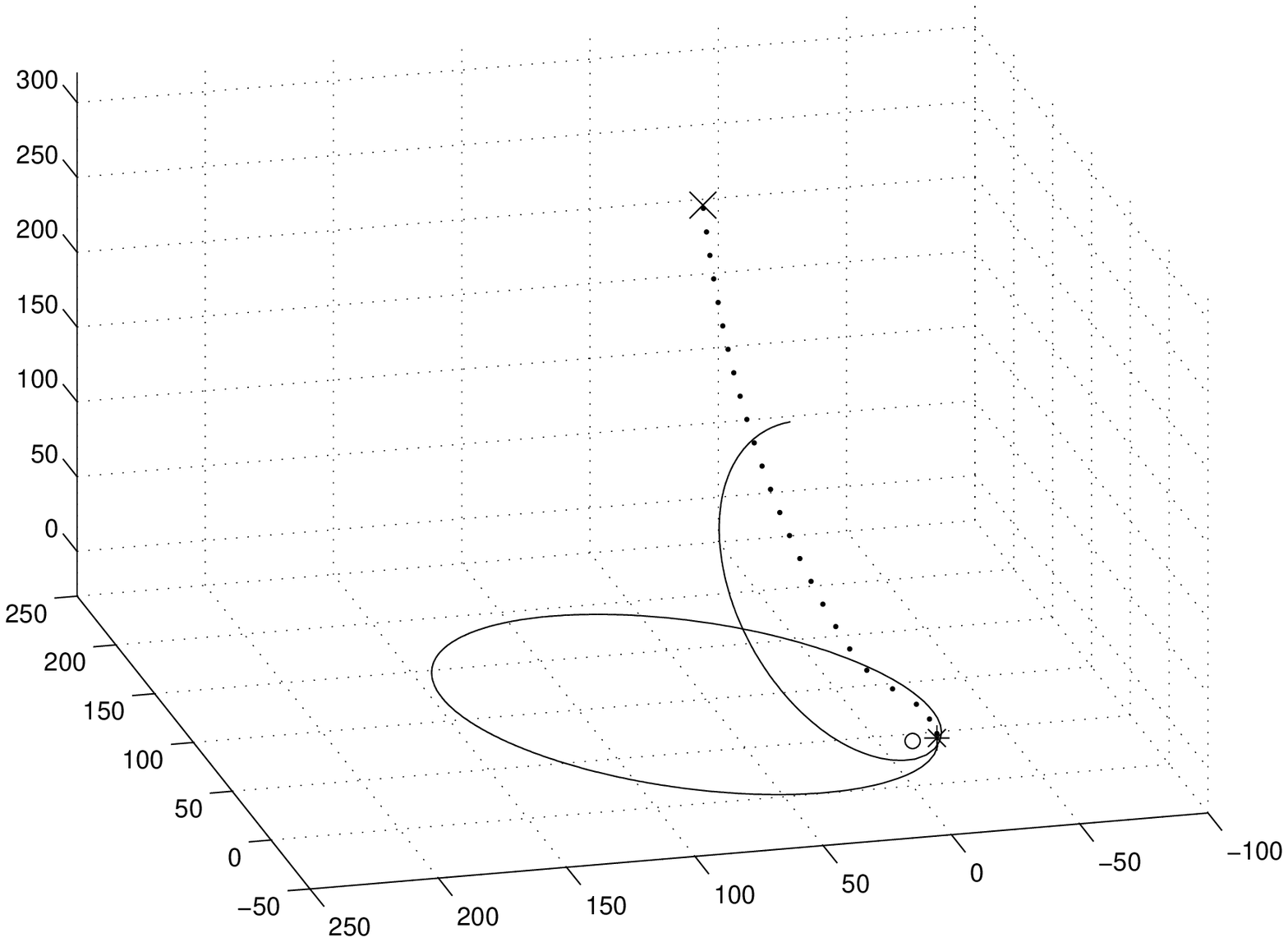}
\includegraphics[scale=0.5]{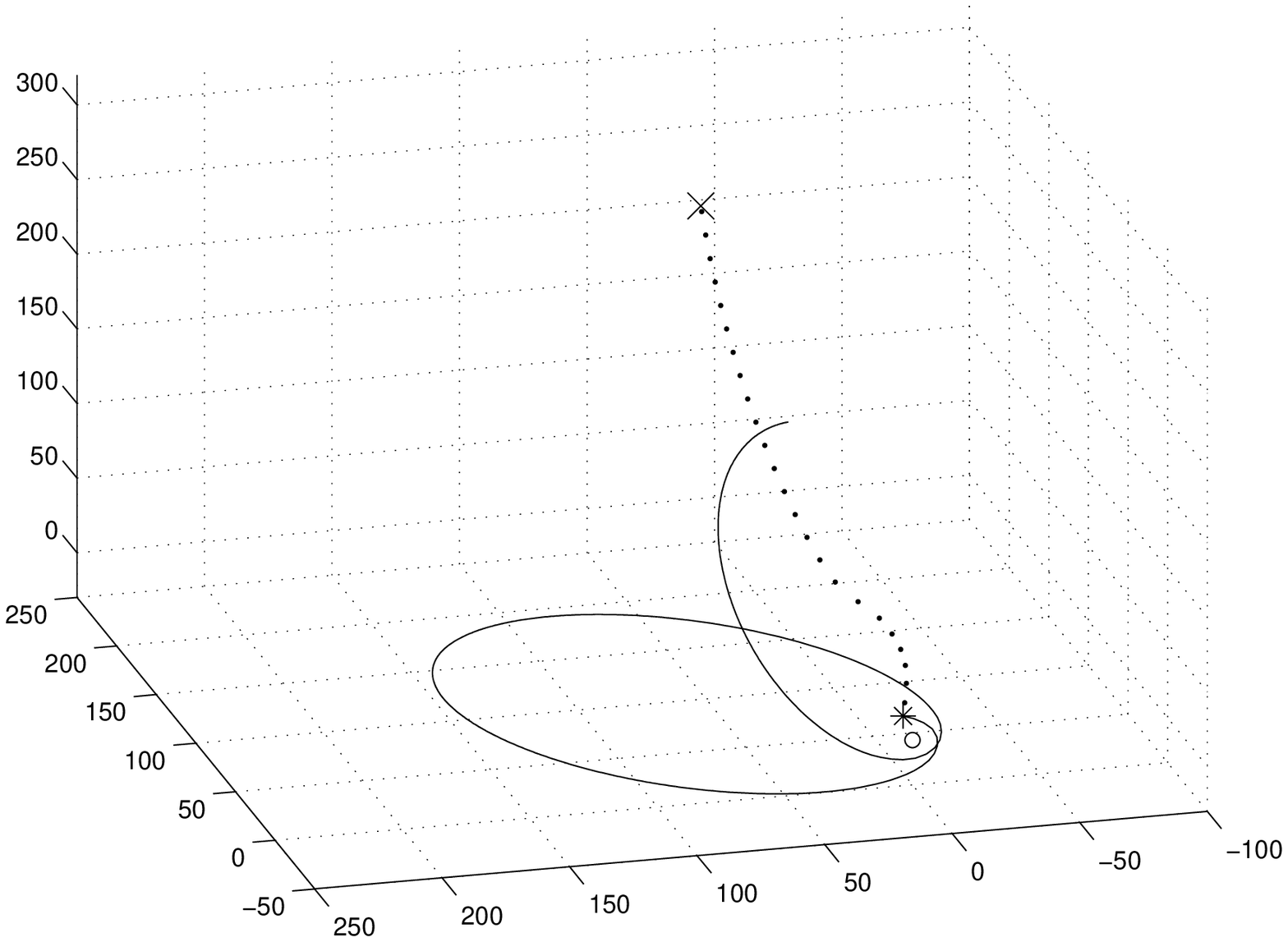}\\
\includegraphics[scale=0.5]{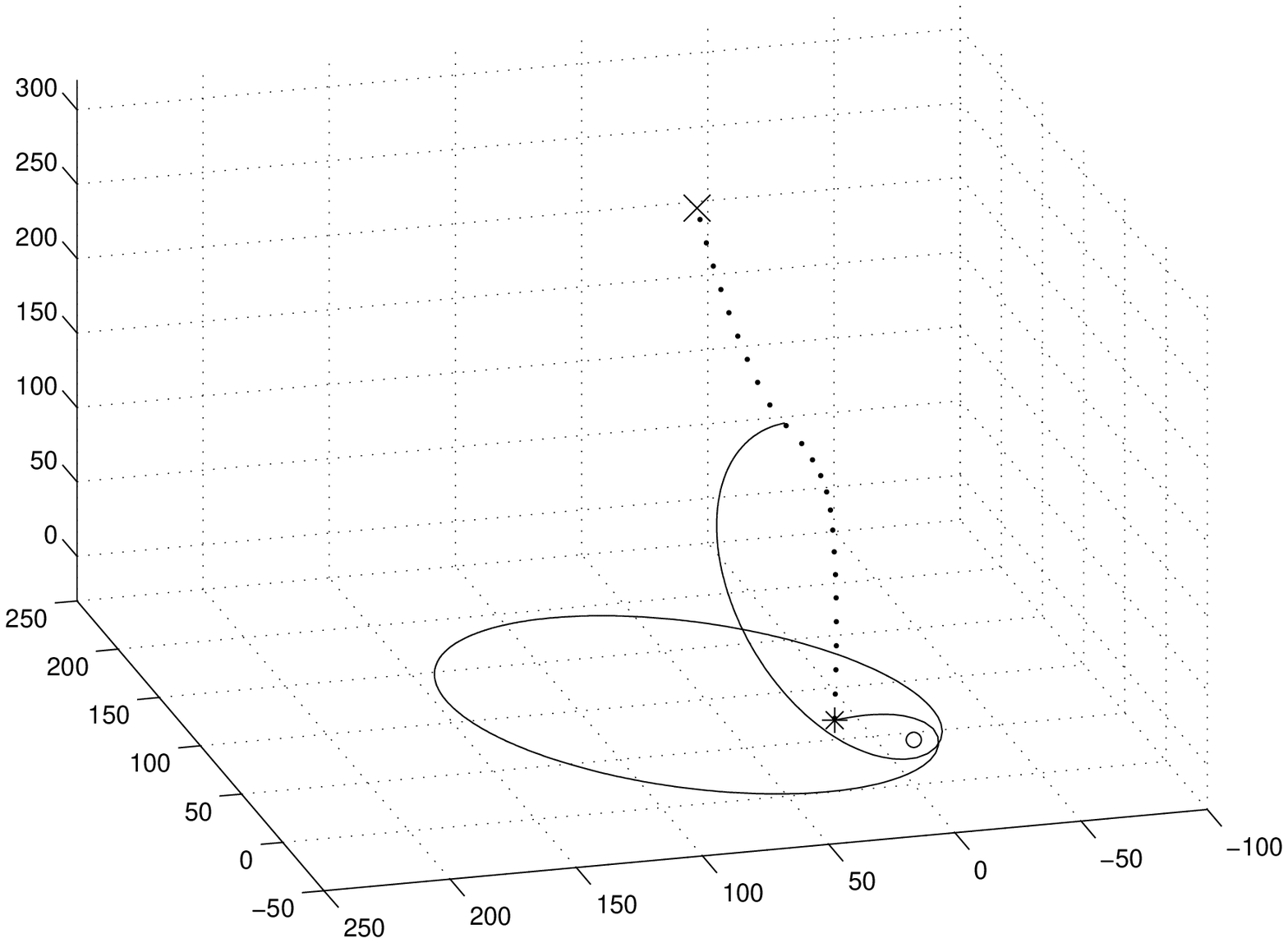}
\includegraphics[scale=0.5]{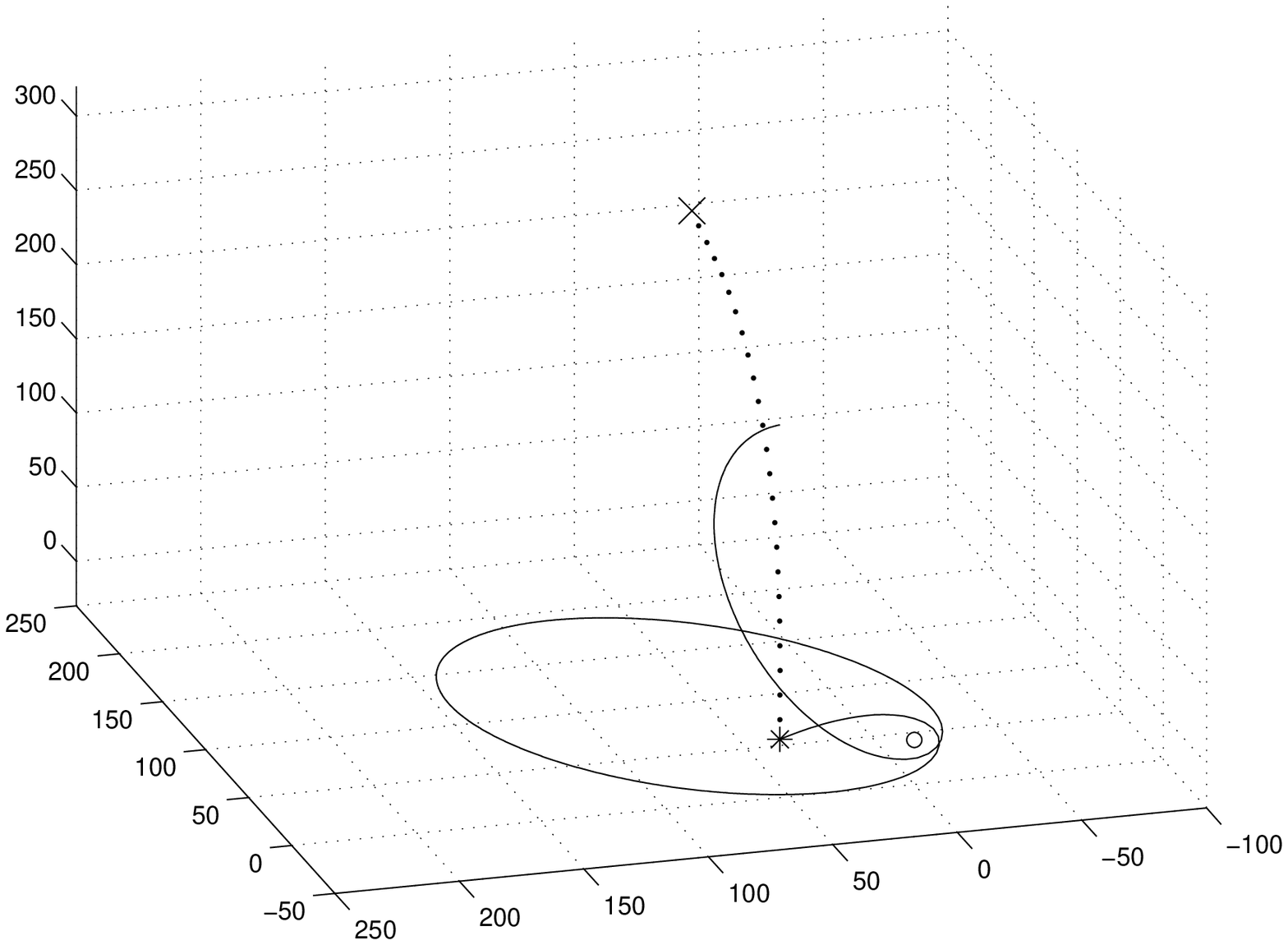}
\caption{Depiction of the system we deal with, shown at four slices of coordinate time.   The curve ending in a star symbol traces a stellar orbit, starting from apocenter, through its second pericenter passage. The star  emits photons in all directions at equal intervals of proper time.  Dots show photons which will reach the observer (marked $\times$).   Note that the dots represent a sequence of photons emitted at different places, hence joining the dots does not represent the path of any particular photon. Time dilation and Schwarzschild perturbations are included in this example, but not frame dragging or quadrupole. This means that the star's orbit precesses, and the photons are lensed. In order to make the relativistic effects visually discernible, the star has a very low value of $a = 100$. Such a star may, however, yet be discoverable by the E-ELT  \citep{EELT}.   Also, the observer has been placed at an unrealistically close distance of 300.}
\label{fig:schematic}
\end{figure}
\end{center}

Several different possibilities for observing relativistic effects
have been discussed in the literature.  To see how these relate to
each other, let us consider the different physical effects involved in
observing a relativistic orbit over time.  These are illustrated in
Figure~\ref{fig:schematic}.  Shown schematically in this figure is a
star on a precessing orbit, which emits photons (pulses, wavecrests)
at a fixed frequency in proper time.  These photons then take a curved
path to the observer.  When they arrive, both their frequency and
their travel direction have been altered.  What can the
observer detecting these photons infer about the metric? This
observer must take multiple physical processes into account.
\begin{enumerate}
\item {\em Time dilation\/} at the source is a consequence of the
  equivalence principle, and is the strongest relativistic
  perturbation. Its measurability in the context of the Global Positioning
  System (GPS) is well known \citep{2003LRR.....6....1A}.  
  The manifestation of time dilation in
  pulsar timing is known as the Einstein time delay.  For the S stars, 
  a spectroscopic detection of time dilation is expected around 
  pericenter passage  \citep{gillessenredshift}.
\item {\em Orbit perturbations\/} are a test of space curvature, and
  at higher order, of black hole spin.
\begin{itemize}
\item The leading order precession of orbital periapse is well measured 
  in binary pulsars. \citep[For a discussion of how pulsar timing effects
  scale to the Galactic-center region, see][.]{PfahlLoeb2004}
  Astrometric measurement of precession in the orbit of S2 is considered
  feasible with current instruments \citep{2009svlt.conf..361E}, but the 
  long orbital period, and the likelihood of Newtonian precession due to the
  distributed mass, both present serious difficulties.  If stars much
  further in are discovered, however, precession of orbital planes due to 
  higher-order spin and quadrupole effects become accessible \citep{2008ApJ...674L..25W,nbody}.
\item Orbital decay due to gravitational radiation is not considered
  measurable because of the long time scales and the extreme mass ratio between the star or pulsar and the supermassive black hole. 
\item In contrast to the above secular effects, there are also
 GR induced velocity perturbations, which vary along the orbit
 \citep{kannan,pretosaha}.  In binary pulsars, velocity perturbations
 are not considered especially interesting and are subsumed within
 Roemer delays. For Galactic center stars however, the situation is
 different. First, the velocity perturbations are larger, and second,
 they are concentrated around pericenter passage, thus offering a
 better prospect for isolating relativistic from Newtonian
 perturbations.
\end{itemize}
\item {\em Light paths and travel times\/} are affected by the black hole mass, and again at higher order, by spin.
\begin{itemize}
\item Strong deflection of starlight \citep{2009ApJ...696..701B} or a
pulsar beam \citep{wang1,wang2} would be a spectacular though rare event.
\item Small perturbations of photon trajectories, too small to measure
astrometrically, can nevertheless produce detectable changes in the
light travel time.  In the pulsar literature this is known as the
Shapiro delay.  The analogous effect for S~stars is a time-dependent
redshift contribution \citep{paper1}, which for a star like S2 is comparable to the
velocity perturbations.  Hence, in the Galactic center, relativistic
perturbations on both orbits and light paths must be considered.
\end{itemize}
\end{enumerate}
Later in this paper, we will group time dilation with the orbital effects.

Motivated by the above considerations, in this paper we develop a
method to fit a relativistic model to observables. This strategy (i)~takes
relativistic perturbations on both the orbits and the light into
account, (ii)~treats stellar spectroscopy and pulse timing in a
unified way, and (iii)~is well suited to analyzing data obtained from
a single orbital period or even less.  We include the effects of time dilation, space curvature, frame dragging, and torquing-like effects induced by the quadrupole moment from the black hole spin. Some further issues remain; most importantly, how to include Newtonian
perturbations (from mass other than the black hole's) in the fit ---
hence the ``towards'' in the title.

The observable we consider is the apparent frequency $\nu$.  As
remarked above, this can refer to spectral lines or pulses.  The
source frequency $\nu_0$ is in principle known for stars, but unknown
for pulsars.  Note that
\begin{equation}
  c\ln(\nu_0/\nu) = c\ln(1+z) \simeq cz
\end{equation}
where $z$ is redshift in the usual definition.  For this paper,
however, we will use ``redshift'' to mean $c\ln(\nu_0/\nu)$.  The
possibly unknown source frequency now appears as a harmless additive
constant in the redshift.  For the S~stars, the current spectroscopic
accuracy is $\approx10\kms$ under optimal conditions
\citep{gillessen}.  If a pulsar on a similar orbit is discovered, the
accuracy would likely be much higher.  Taking, as an example, a
one-hour observation with pulses timed to $3\,\mu\rm s$
\citep[cf.][]{milliseconds} implies an accuracy of one part in $10^9$,
equivalent to a redshift accuracy of $30\cms$.  The same level of
accuracy is not inconceivable from spectroscopy of S~stars, since
planet searches routinely achieve $\Delta v<1\ms$ 
\citep[e.g.][]{2006SPIE.6269E..23L},
but would require some technical breakthroughs to achieve for faint
infrared sources like the S~stars.

For a given orbit and redshift accuracy, the principal quantities we will calculate are the signal-to-noise ratio $S/N$
for different relativistic effects. For each of four relativistic effects --- time
dilation, space curvature, frame dragging, and quadrupole moment effects --- we will
present the results in two ways: (i)~the redshift accuracy needed to
reach some $S/N$ for a given orbit, and (ii)~the orbital parameters 
needed to reach some $S/N$ for a given redshift accuracy.

\section{Model}

The main calculations in this paper will assume that the trajectories of both
S~stars (or pulsars) and photons are geodesic in a pure Kerr metric. Naturally, geodesics of the former are timelike, and the latter null. Even if Einstein gravity is correct, the pure Kerr metric is naturally an approximation because this solution to the field equations neglects all mass in the vicinity of the black hole (more on this later), and the mass of the star itself. In general relativity, geodesic
motion is fully described by the super Hamiltonian
\begin{equation}
\H = \half \, g^{\mu\nu}p_\mu p_\nu,
\end{equation}
with $\H=0$ for null geodesics.

Rather than carry out all computations with the Hamiltonian resulting from the full Kerr metric, it is useful to consider two different
approximations for the cases of orbits and light paths. These we will
call $H_\star$ and $H^\photon$.  (Word labels in subscripts indicate
orbits, in superscripts light paths.)  We express these approximate
Hamiltonians perturbatively as 
\begin{equation}\label{Hstar}
\H_\star = \H_\static + \epsilon^2\, \H_\kep + \epsilon^4\, \H_\schw
           + \epsilon^5\, \H_\fd + \epsilon^6\, \H _\torq,
\end{equation}
and
\begin{equation}\label{Hnull}
\H^\photon = \H^\mink + \epsilon^4\, \H^\schw
           + \epsilon^6 \left( \H^\fd \, + \H^\torq \right),
\end{equation}
where the meanings of the various component Hamiltonians will be
explained shortly. 
Note that $\epsilon$ is just a label to indicate
orders: if $\epsilon^n$ appears, the term is of order $v^n$ (where $v$
is the stellar velocity in light units), but numerically $\epsilon=1$.
The series in (\ref{Hstar}) and (\ref{Hnull}) actually consist of the
same terms, but because $\H_\star$ is to be applied only to
trajectories that are strongly timelike, and $\H^\photon$ only to
compute null geodesics, the orders of same terms are often different.

Expressions for all the Hamiltonian terms are derived perturbatively
from the Kerr metric, using the basic method explained in
\cite{paper1}.  We will not repeat the details here, but simply list
the terms and their physical meanings.

First we consider the orbit (Equation \ref{Hstar}).

\begin{itemize}
\item To leading order, the star feels no gravity and the clock simply ticks:
\begin{equation}
\H_\static = -\frac{p_t^2}{2}.
\end{equation}
\item At next-to-leading order, gravity first appears, with
\begin{equation}\label{Hkep}
\H_\kep = \frac{p_r^2}{2} + \frac{p_\theta^2}{2r^2}+ 
\frac{p_\phi^2}{2r^2\sin^2\theta}-\frac{p_t^2}{r},  
\end{equation}
which is the classical Hamiltonian modified by $1/r\rightarrow
p_t^2/r$. This modification leaves the problem spatially
unchanged, but introduces a temporal stretching causing a
gravitational time dilation. This is originally a consequence of the
Einstein Equivalence Principle, and affects the redshift at $\O(v^2)$.
\item Appearing next are the leading order Schwarzschild contributions,
\begin{equation}
\H_\schw = - \frac{2p_t^2}{r^2}-\frac{p_r^2}{r},
\end{equation}
of which the first term effects a further time dilation, and the
second curves space, leading to orbital precession.
\item Next, the first spin term emerges with
\begin{equation}
\H_\fd = - \frac{2s p_t p_\phi}{r^3}.
\end{equation}
producing frame dragging, an $r$-dependent precession around the spin
axis.  Here $s$ is the spin parameter, which points in the $\theta = 0$ or $+z$
direction. Maximal spin is $s=1$.
\item The last set of orbital terms we will consider are
\begin{equation}
\H_\torq = s^2\left(\frac{\sin^2\theta p_r^2}{2r^2} -\frac{p_\phi^2}{2r^4\sin^2\theta} + \frac{\cos^2\theta p_t^2}{r^3} - \frac{\cos^2\theta p_\theta^2}{2r^4}\right),
\end{equation}
which are the leading-order quadrupole moment terms.  These result in the
orbit being torqued towards the spin-plane, as well as other less
intuitive effects.

\end{itemize}

Then we consider light paths (Equation \ref{Hnull}).
\begin{itemize}
\item At leading order, we have
\begin{equation}
\H^\mink =-\frac{p_t^2}{2}+\frac{p_r^2}{2}+\frac{p_\theta^2}{2r^2}+
\frac{p_\phi^2}{2r^2\sin^2\theta},
\end{equation}
which is to say, the spacetime is Minkowski and the photon
trajectories are straight lines.
\item The leading order Schwarzschild
contribution
\begin{equation}
\H^\schw = -\frac{p_t^2}{r}-\frac{p_r^2}{r}
\end{equation}
(which notice is not identical to $\H_\schw$) introduces lensing and
time delays.
\item The frame-dragging term for photons
\begin{equation}
\H^\fd =  - \frac{2s p_t p_\phi}{r^3}
\end{equation}
debuts at one higher order than $\H_\fd$.  Frame dragging lenses photons around the black hole in a twisted fashion. A
$\phi$-dependent time-dilation also occurs.
\item Finally, we consider higher order Schwarzschild terms, and
  quadrupole moment terms
\begin{equation}
\H^\torq = -\frac{2p_t^2}{r^2} + \frac{s^2\sin^2\theta p_r^2}{2r^2}
- \frac{s^2\cos^2\theta p_\theta^2}{2r^4}
- \frac{s^2p^2_\phi}{2r^4\sin^2\theta}
\end{equation}
which, as before, produce a torque towards the spin plane and as well
as some more subtle effects.
\end{itemize}

The well-known separability
properties of the Kerr metric \citep[cf.][]{1983mtbh.book.....C}
provide solutions up to quadratures, but not explicit solutions.
Analytic solutions for geodesics and null geodesics are available for
various special cases, for example for the leading-order Schwarzschild
case \citep{dando}. But the present work uses numerical integration of
Hamilton's equations.

\section{Calculating redshifts}

Given a set of orbital parameters for the star, we calculate a
redshift curve, meaning $\ln\left(\nu_0/\nu\right)$ against observer
time or pulse arrival time, by the following method.

First the orbit is calculated by numerically solving Hamilton's
equations for $\H_\star$ (Equation \ref{Hstar}).  If desired,
particular orbital relativistic effects can be isolated by omitting other
relativistic terms from $\H_\star$.  The independent variable is, of
course, not time but the affine parameter (say $\lambda$).  Along this
orbit, from pairs of points separated by $\Delta\lambda$, photons are
sent to the observer.  The difference in proper time between the
emission of two photons is
\begin{equation}
   \Delta\tau = \frac{\Delta\lambda}{\sqrt{-2\,\H_\star}}.
\end{equation}
The photons themselves travel along paths determined by $\H^\photon$
(Equation \ref{Hnull}) with the additional condition $\H^\photon=0$.
Again, one is free to isolate particular relativistic effects on the
photons by discarding terms from $\H^\photon$.  If $\Delta t$ is the
difference in the photons' times of arrival at the observer, then
\begin{equation}\label{problem}
\frac{\nu_0}\nu = \frac{\Delta t}{\Delta \tau}.
\end{equation}
This calculation is carried out along the orbital path. Some snapshots of this procedure are illustrated in Figure~\ref{fig:schematic}.

Now, photons are emitted from the star in all directions, but we need
to find exactly those photons which reach the observer. Solving this
boundary value problem is the computationally intensive part.  Our
algorithm for doing so is explained in \cite{paper1}.  In stronger
fields, the redshift curve takes longer to compute than in weaker
ones, both because the orbit integration needs smaller stepsizes over
more relativistic regions, and because the boundary value problem 
requires more iterations before satisfactory convergence is reached.

\begin{figure}
\includegraphics[scale=0.93]{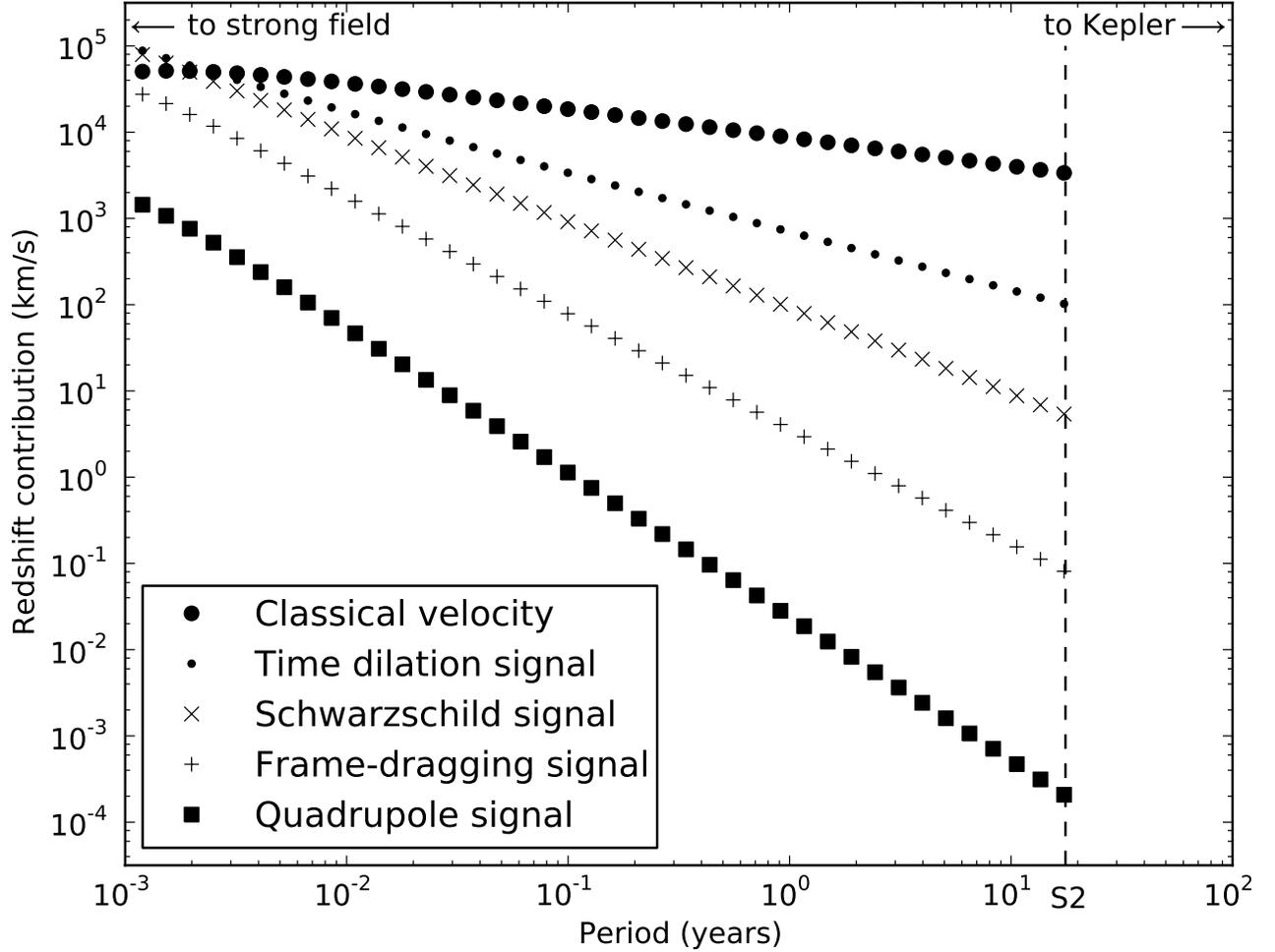}
\caption{Redshift contributions of different orbital effects.  Shown
  here is the contribution of each term in Equation (\ref{Hstar}) at
  pericenter, with the orbital period being varied and the other
  orbital parameters fixed at the S2 values.  The pericenter distance
  varies from 3600 (approximately the value for S2) down to 6. Note that for the last two signals - the frame-dragging and the quadropole - we have taken the black hole spin to be maximal 
and to point perpendicular to the line of sight. The signal scalings are listed in Table \ref{effect_grouping}. The curves begin to intersect as we move towards the strong field regime - attributable to the the breakdown in our perturbative approximation for small $r$.  }
\label{fig:orbital_signals}
\end{figure}

\begin{figure}
\includegraphics[scale=0.93]{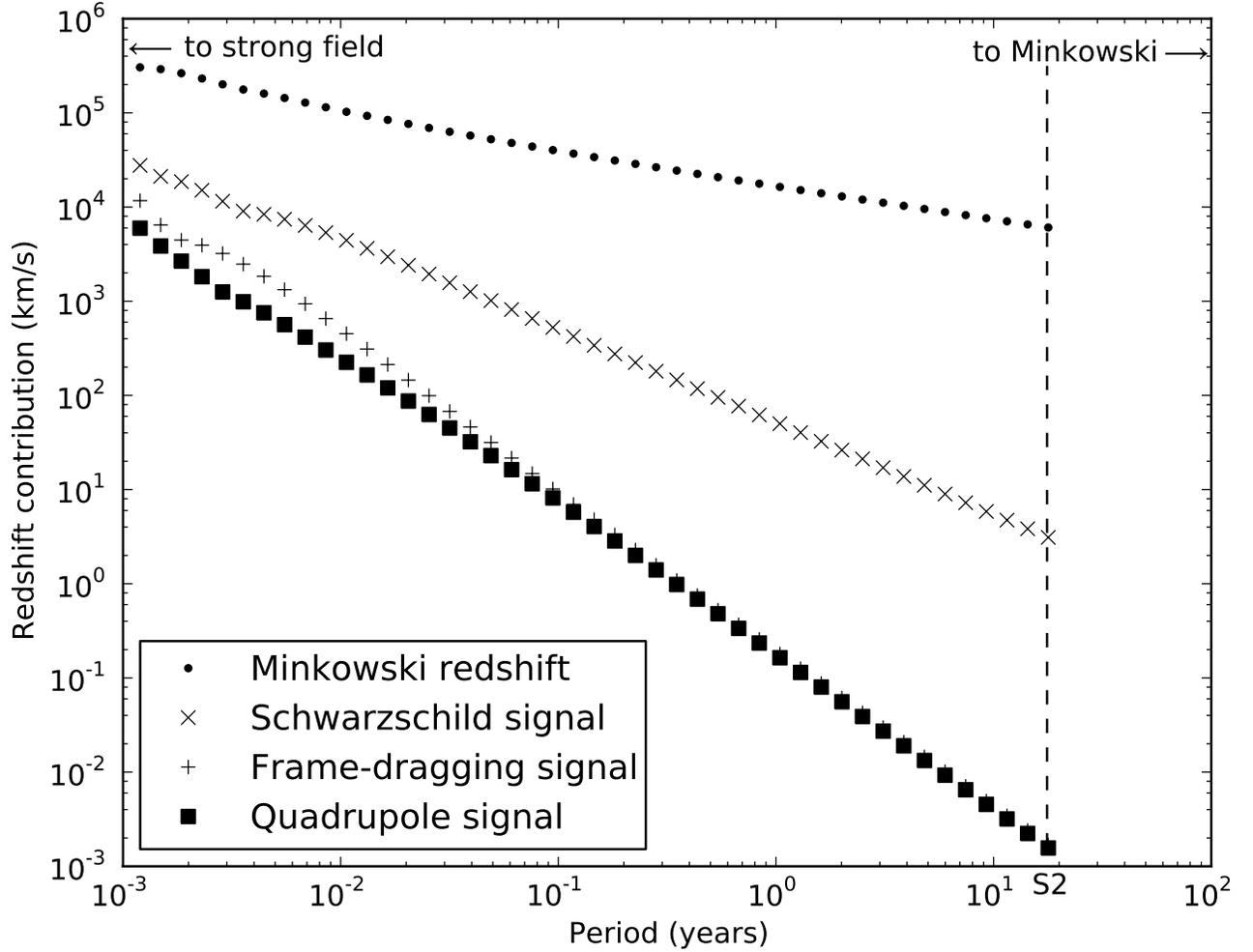}
\caption{Like Figure~\ref{fig:orbital_signals} but for photon propagation
  delays (Equation \ref{Hnull}).  The Schwarzschild propagation signal is for the most part slightly smaller than the Schwarzschild orbital signal, and scales in the same way. The frame-dragging propagation signal is considerably smaller than the corresponding orbital signal, and scales like $z^{\fd} \propto P^{-5/3}$, as opposed to $z_{\fd} \propto P^{-4/3}$. The frame-dragging signal remains approximately an order of magnitude weaker on the photons than on the star. This suggests that in attempting to measure the spin of the black hole in the post-Newtonian regime, its manifestation on the orbit is what matters. Note however, that this is not the case for the Schwarzschild effects - for which neither the photon nor orbit perturbations may be neglected. }
\label{fig:photon_signals}
\end{figure}

The observable redshift is of course a combination of all the
relativistic effects atop the classical redshift.  One can, however,
estimate the strengths of each effect in isolation, by toggling each
term in Equations (\ref{Hstar}) and (\ref{Hnull}) on and off, and then
taking the difference in redshift.  Figures \ref{fig:orbital_signals}
and \ref{fig:photon_signals} show relativistic redshift contributions
measured in this way.  These orbits refer to the pericenters of a
range of orbits, varying in $a$ but with $e$ and the other orbital
parameters fixed at the values of S2.  
These effects have simple
scalings with the orbital period, readily deduced from the scaling
properties of the Hamiltonians, and summarized in
Table~\ref{effect_grouping}.  Numerical results in \cite{paper1}
verify that such scalings apply not only at pericenter but all along
the orbit.

\begin{figure}[!h]
\includegraphics[scale=0.93]{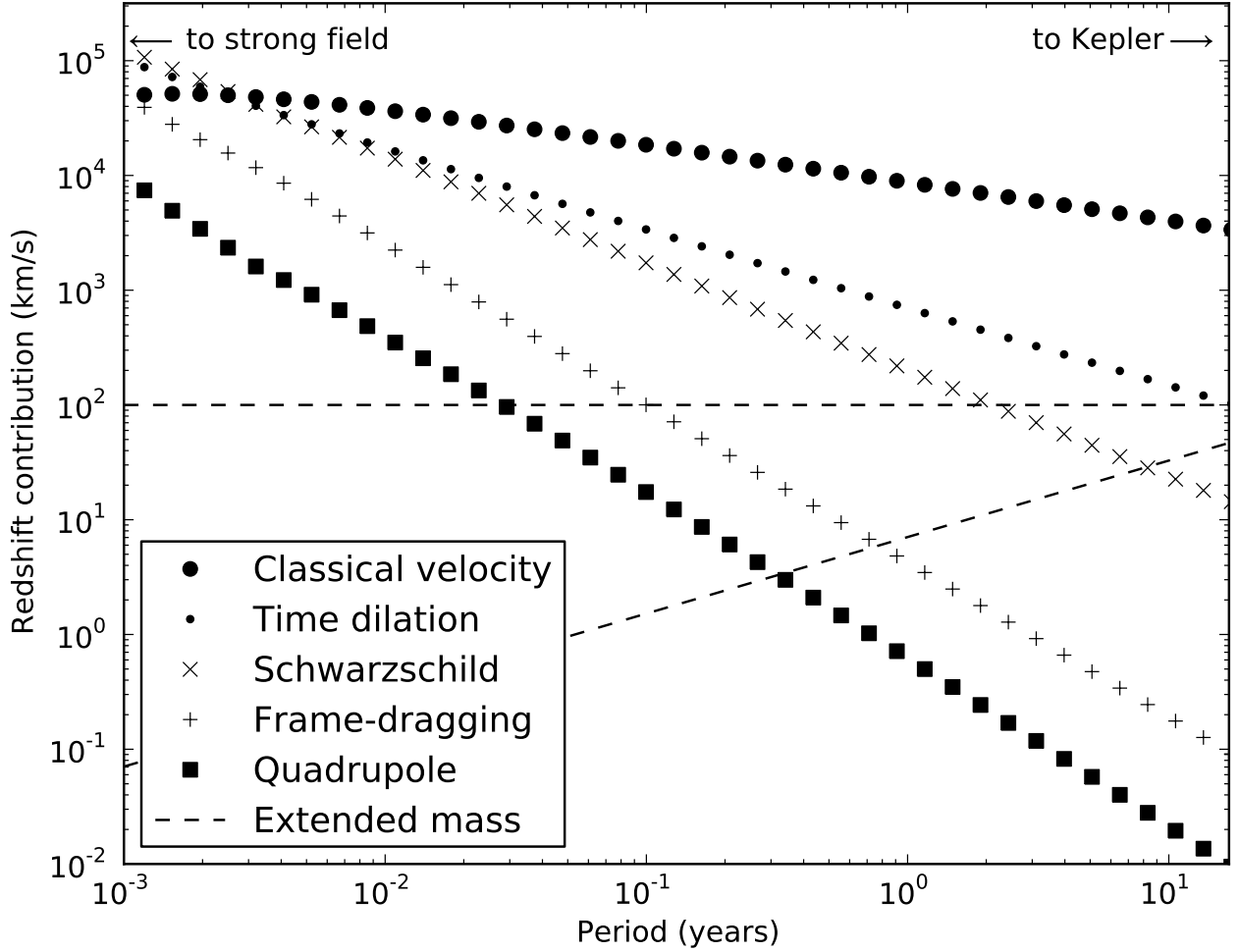}
\label{fig:extended_mass.eps}
\caption{Relativistic redshift effects, with orbital and light-path
  contributions summed.  Two estimates for the signal due to the
  extended mass distribution are also shown, using the crude model
  (\ref{sphericalmodel}) normalized so that the circular velocity at
  the $r=10^5\,G\Mbh/c^2$ (or $\sim 0.1\,\textrm{pc}$) is  $\sim 100\kms$ \citep[cf.][]{gillessen}.  
  The flat and sloping dashed curves correspond to $\gamma=2.5$ and 1.5
  respectively.}
\label{fig:sum}
\end{figure}

The extended mass distribution in the Galactic center region,
due to all the other stars, stellar remnants and dark matter
particles that are present, introduces Newtonian perturbations.  
The distribution and
normalization of this mass are poorly constrained on the scales
of interest \citep{2009A&A...502...91S}, 
although some upper limits exist \citep{2009ApJ...707L.114G}
and some numerical experiments have been done \citep{nbody}.  
Assuming spherical symmetry for this distributed mass,
models of the form
\begin{equation}
\rho\left(r\right) \propto r^{-\gamma}
\label{sphericalmodel}
\end{equation}
are sometimes adopted. 
Such a model ignores the torques due a nonspherical
or discrete mass distribution, and the only change it implies
in the orbital dynamics is an additional (prograde) pericenter advance.
Figure \ref{fig:extended_mass.eps} compares the orbital 
and photon signals strengths for each effect, 
including an extended mass distribution of the type (\ref{sphericalmodel}). Note however, that in the analysis which follows, we do not attempt to include the effects of the extended mass distribution. 

\section{Fitting}

The computational demands for the present work are far
greater than calculations of a few redshift curves for given parameters
\citep[as in][]{paper1}, because {\em fitting\/} redshift curves in a
multidimensional parameter space requires evaluations of up to tens of
thousands of such curves.  To this end, parallel functionality was
added to the implementation. The work is distributed evenly among the
available processors.  A workload is the charge of calculating the
redshift from a single position on the star's orbit, i.e., a single
evaluation of (\ref{problem}), calling for the finding of two photons
travelling from star to observer. The program is available as an
online supplement.

\subsection{The parameters}

The redshift curve of an S~star or pulsar depends on eight essential
parameters, but there can be any number of additional parameters for
secondary effects.  In this paper we consider nine parameters in all,
as follows.

\begin{itemize}
\item The black hole mass $\Mbh$ sets the overall time scale, and
  accordingly we express it as a time.  Changing $\Mbh$ simply
  stretches or shrinks the redshift curve in the horizontal direction.
\item The intrinsic frequency $\nu_{0}$ for pulsars is the pulse
  frequency in proper time, whereas for spectroscopy, it has the
  interpretation of absolute calibration.  Altering $\nu_0$ simply
  shifts the redshift curve vertically.
\item Then there are the Keplerian elements, referring to the
  instantaneous pure Keplerian orbit with the initial coordinates and
  momenta.  Our orbit integrations all start at apocenter.
\begin{enumerate}
\item The semi-major axis $a$ (or equivalently the period
  $P=2\pi a^{3/2}\Mbh$) is single most important parameter dictating
  the strength of the relativistic signals.
\item The eccentricity $e$ sets how strongly peaked the redshift curve
  is at pericenter.
\item The argument of pericenter $\omega$ sets the level of asymmetry
  of the redshift curve.
\item The orbital inclination $I$ changes the amplitude of the
  redshift.  Classically, the inclination enter the redshift depends
  only on $\Mbh\sin^3 I$ (on $\Mbh^3/M_{\rm total}^2\sin^3 I$ for
  finite mass ratio).  In relativity, the degeneracy is broken,
  because time dilation is independent of $I$.
\item The epoch $\epsilon_0$ basically the zero of the observer's clock.
\end{enumerate}
The longitude of the ascending node $\Omega$ does not appear, because,
in our chosen coordinate system, $\Omega$ rotates the whole
system about the line of sight, which has no effect on the redshift.
Formally, we simply fix $\Omega = 0$.

\item For the pulsar case, we include a simple spin-down model, with a
  constant spin-down rate $\nu_0 \rightarrow \nu_{0} -
  \dot{\nu}\left(t_a\right)$.  Here $\dot\nu$ is an additional
  parameter.
\end{itemize}

We take the black hole spin as maximal, and to point perpendicular to
the line of sight.  Proper motion of the black hole is not considered.

To fit, we use a quasi-Newton limited-memory Broyden-Fletcher-Goldfarb-Shanno
optimization routine with bounds \citep{lbfgsb, lbfgsb2}.

\subsection{Signal to noise}

With an orbit fitting algorithm in hand, we now need to quantify the
notion of detectability of particular relativistic effects.  One way
would be to attach a coefficient to each term in the relativistic
Hamiltonians and then see how accurately that coefficient can be
recovered. For this paper, however, we take a simpler approach.  Since, for now,
we only aim to identify which effects and which regimes are promising,
we will simply attempt to estimate the threshold for detecting the
presence of each term in the Hamiltonians (\ref{Hstar}) and
(\ref{Hnull}).  Accordingly, we proceed as follows.
\begin{enumerate}
\item For some chosen parameters, we generate a redshift curve
including all Hamiltonian terms up to some $\O(\epsilon^n)$.  These
redshift curves are sampled at 200 points, with dense sampling near
pericenter.
\item We add Gaussian noise to the redshifts, at some chosen level.
\item We then fit the parameters, with a redshift curve {\em
lacking\/} a particular Hamiltonian contribution.  If the noise
level is too high, a reduced $\chi^2\simeq1$ will be obtained,
otherwise $\chi^2_{\rm red}$ will be higher.  We define
\begin{equation}\label{SNR}
S/N \equiv \sqrt{\chi^2_{\rm red}}
\end{equation}
as the signal to noise.
\end{enumerate}
Naturally, it is necessary to check that a large $\chi^2_{\rm red}$ is
not the result of some algorithmic problem, by verifying that a good
fit is obtained if the Hamiltonian term in question is consistently
included.

\subsection{Redshift-accuracy demands for S2}

\begin{figure}[!H]
\includegraphics[scale=0.93]{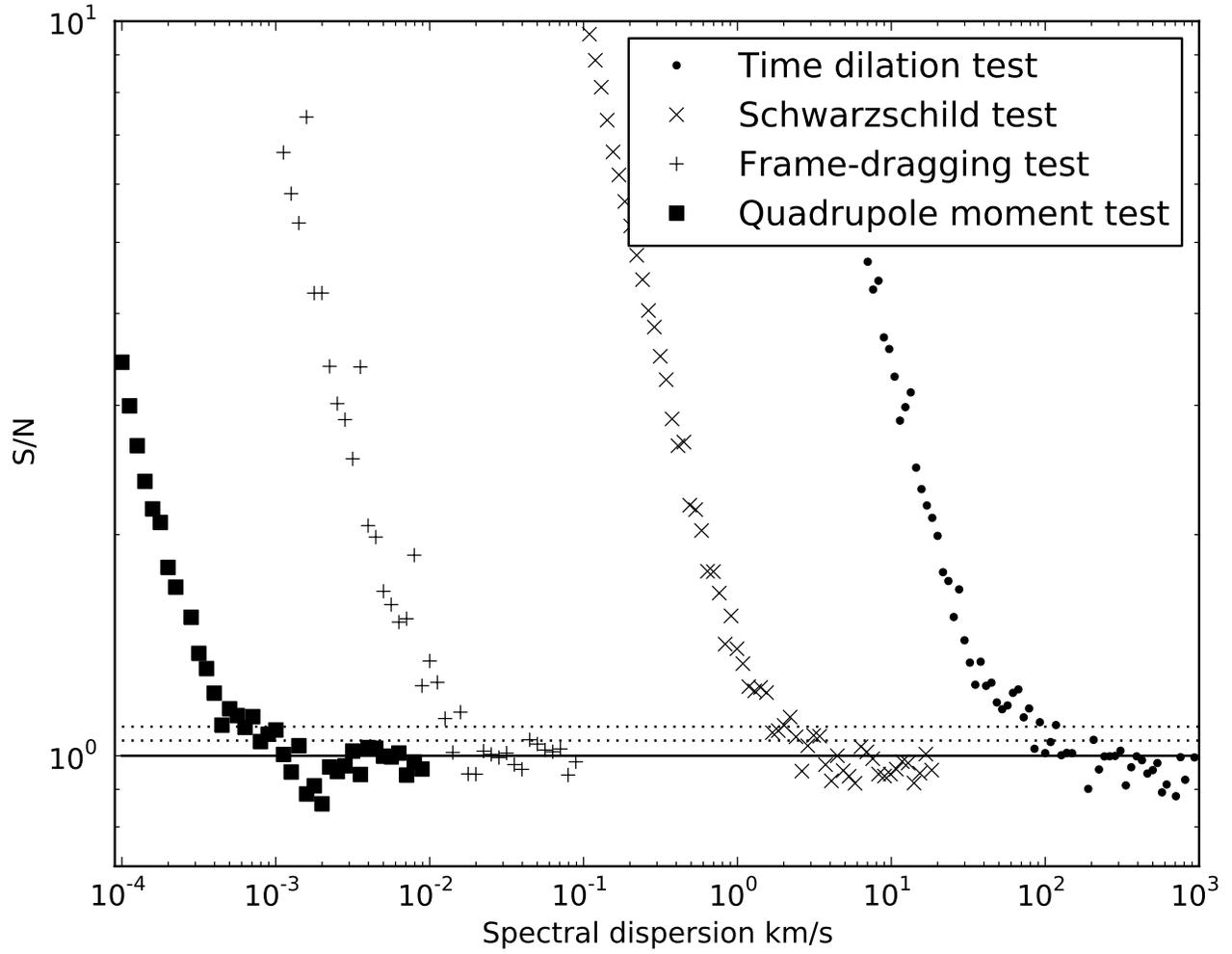}
\caption{The signal-to-noise ratio (as defined by equation \ref{SNR}) for different relativistic effects
  in the orbit of S2 as a function of redshift accuracy.  Each point
  on this figure corresponds to a simulated data set of 200 redshifts,
  distributed over the orbit but with highest density around
  pericenter.}\label{fig:at_30000}
\end{figure}

Figure~\ref{fig:at_30000} shows the $S/N$ of time dilation, space
curvature, frame dragging, and quadrupole terms, all as a function of
redshift accuracy, assuming the orbit of S2.  Time dilation is well
above the current spectroscopic threshold of $\approx 10\kms$.  Space
curvature is somewhat below, while frame dragging and quadrupole are
far below.

The detectability thresholds are summarized in the first part of
Table~\ref{results_summarised}.

\subsection{Orbit-size demands for given redshift accuracy}

\begin{figure}[!H]
\includegraphics[scale=0.93]{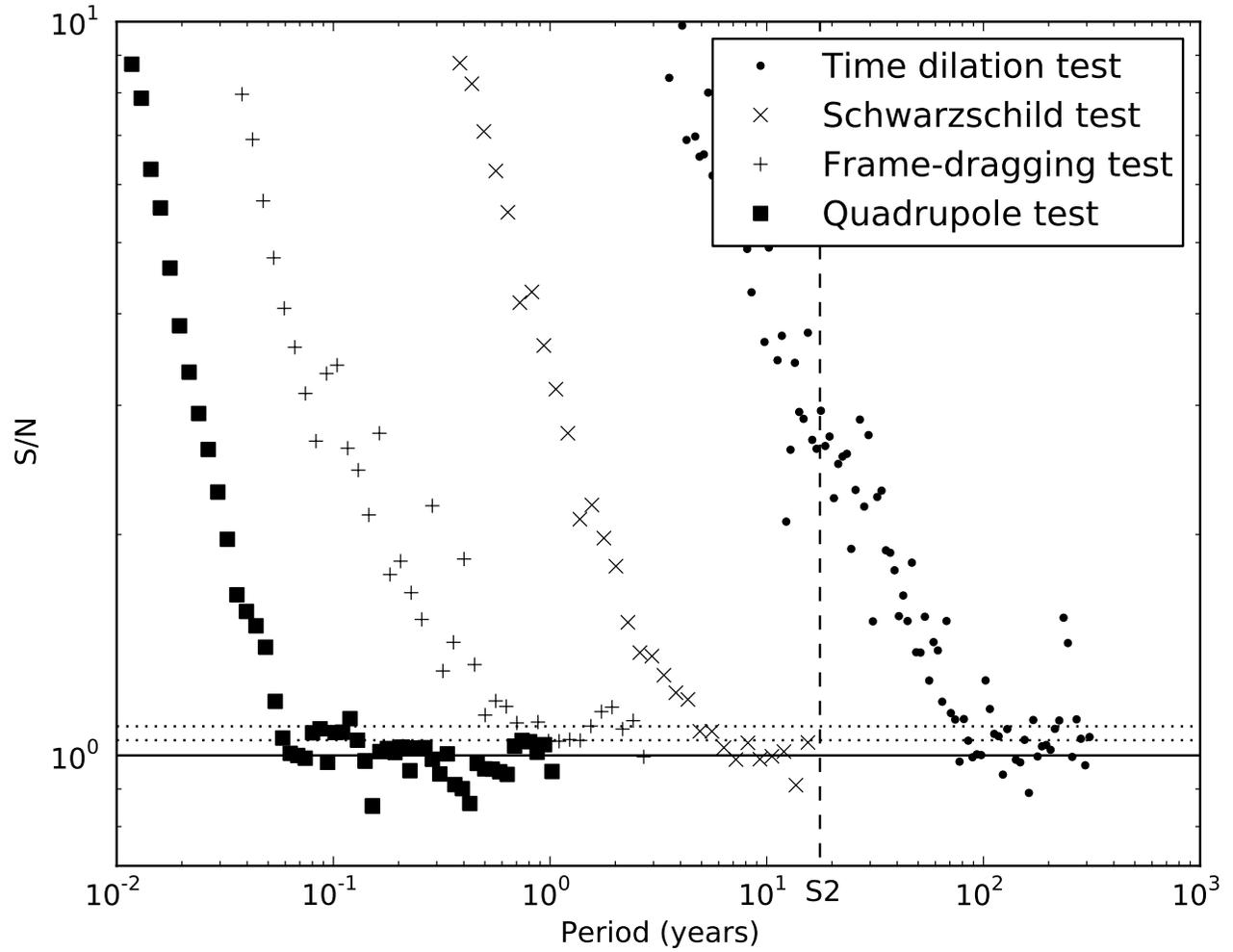}
\caption{Similar to Figure \ref{fig:at_30000}, except that the redshift
  accuracy is fixed at $10\kms$ and the orbits vary, being scaled-down
  and speeded-up versions of the orbit S2. For large $S/N$ these
  curves scale with the powers in Table~\ref{effect_grouping}. Current instrumentation, operating under optimum conditions, should manage an accuracy of $10\kms$ indicating that gravitational time dilation should be able to be detected on some of the currently known S~Stars. }
\label{fig:at_acc_10}
\end{figure}

\begin{figure}[!h]
\includegraphics[scale=0.93]{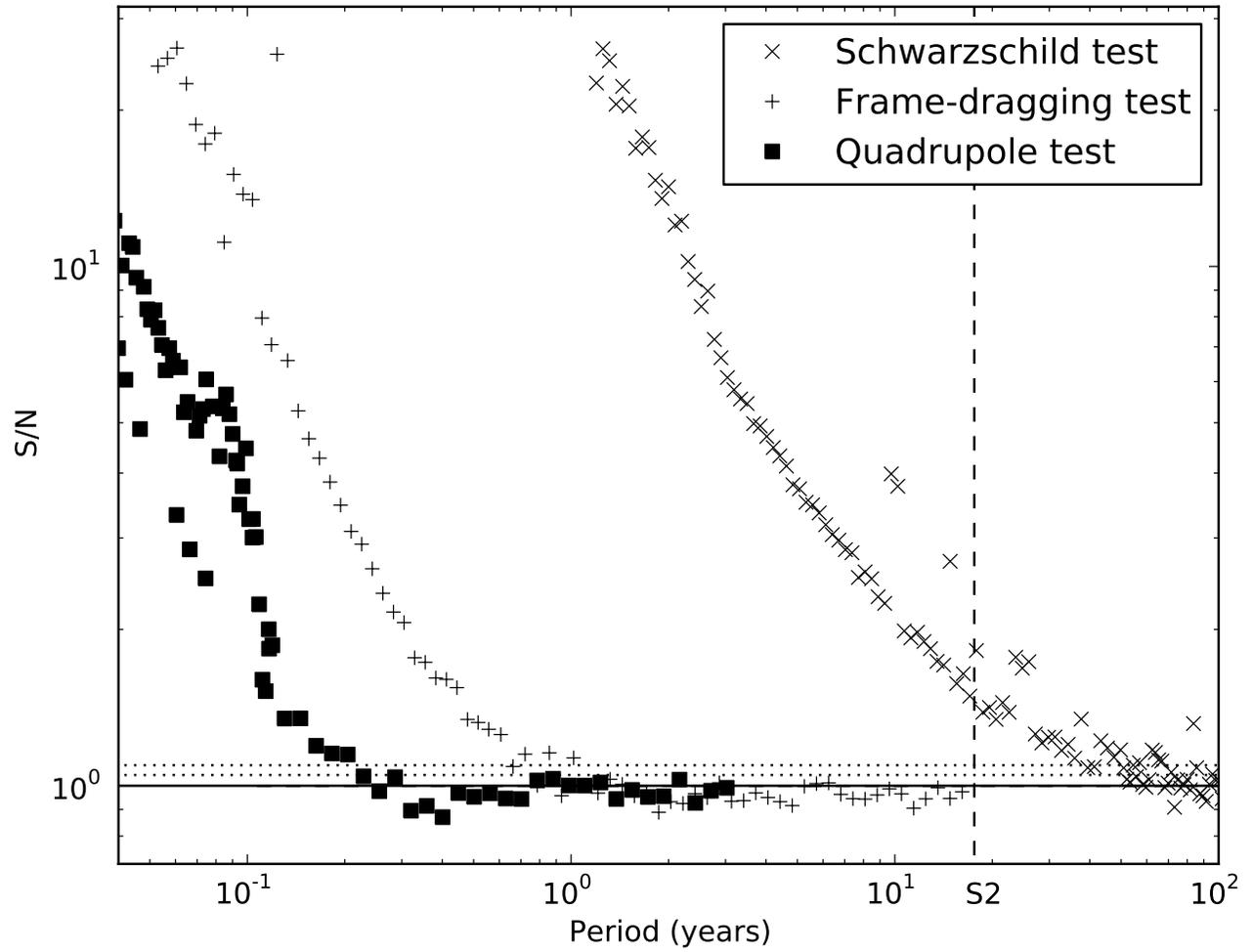}
\caption{Like Figure~\ref{fig:at_acc_10} but for a redshift accuracy
  of $1\kms$ - matching the capabilities of the E-ELT \citep{EELT}.}
\label{fig:at_acc_1}
\end{figure}

\begin{figure}[!h]
\includegraphics[scale=0.93]{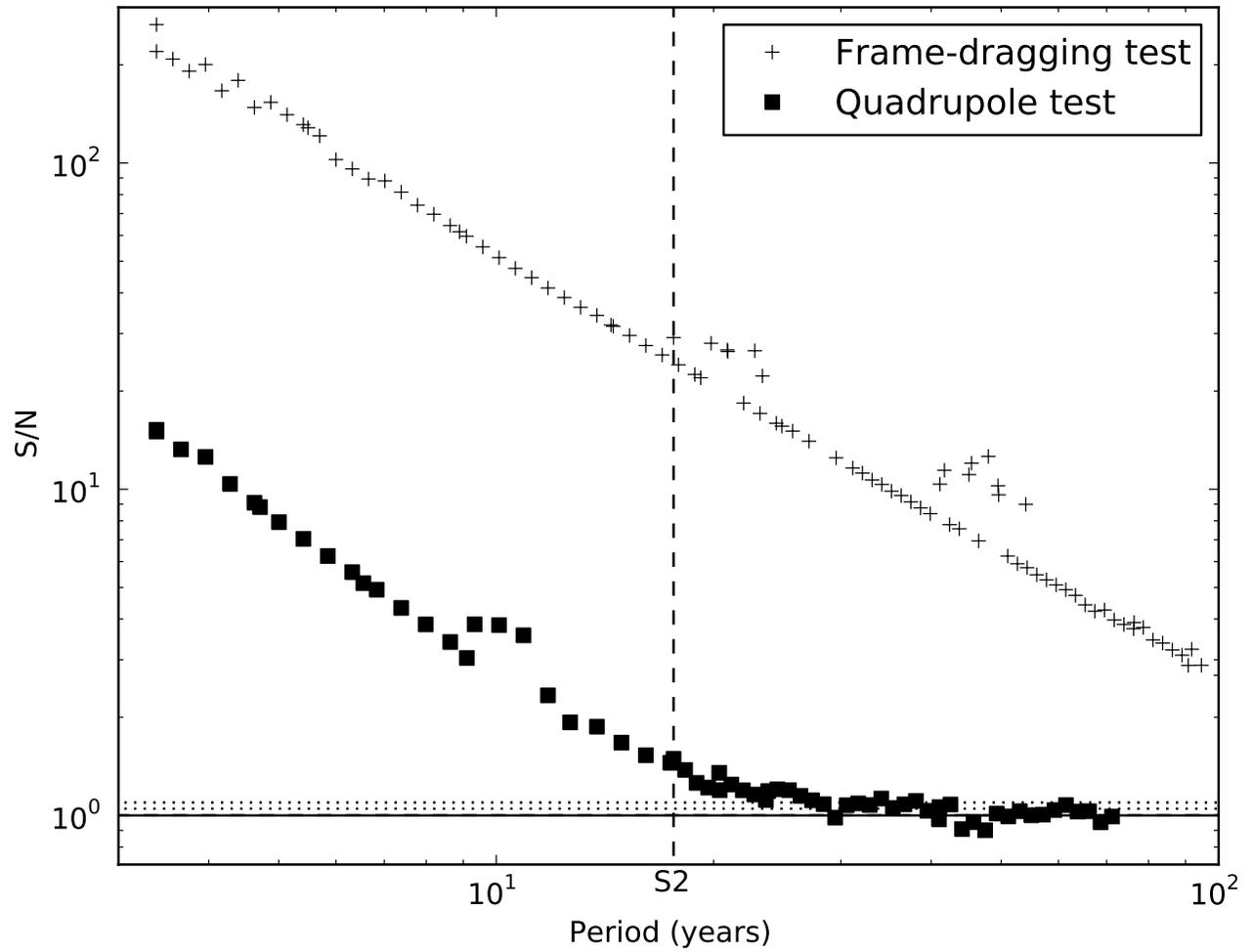}
\caption{Like Figures~\ref{fig:at_acc_10} and \ref{fig:at_acc_1} but
  for redshift accuracy $30\cms$. Pulse timing accuracies at this level are already available, known pulsars orbiting the black hole on suitably short orbits however are not.}
\label{fig:at_acc_3}
\end{figure}

In Figures~\ref{fig:at_acc_10}, \ref{fig:at_acc_1} and
\ref{fig:at_acc_3}, the redshift accuracy is set to $10\kms,$ $1\kms$
and $30\cms$ respectively, while the orbital period varies along the
horizontal axis.

Table~\ref{results_summarised} summarizes the detectability thresholds.

\section{Summary and Outlook}

This paper addresses the problem shown schematically in
Figure~\ref{fig:schematic}, which is to infer orbital parameters and
detect relativistic effects from redshifts or pulsar timings along an
orbit.  The specific observable of interest is the ratio of rest-frame
to observed frequency $\nu_0/\nu$ and how it varies along an orbit,
especially around pericenter.  It is not necessary to measure $\nu_0$
separately, it can be treated as a parameter to be inferred.  As a
result, spectroscopic redshifts and pulsar timings can be treated in a
unified way.  We argue that redshift variation over one or a few
orbits, with special attention given to pericenter passage,
provides a possible route to testing relativity using Galactic-center
stars, or (if discovered) pulsars on similar orbits.  
There are two reasons why a different strategy is called for here
than in the binary pulsar case. 
First, pericenter speeds of S~stars are typically much
higher than those of binary pulsars, and second, 
the orbital periods are too long
for cumulative effects to build up. 

The observable redshift contains several different relativistic
contributions affecting the orbit of the star or pulsar and the
light traveling to the observer.  For calculations, we use a
four-dimensional perturbative Hamiltonian formulation derived from the
Kerr metric.  Each relativistic effect appears conveniently as a
Hamiltonian term that can be toggled on and off to examine its import.
Computation of redshift curves using this Hamiltonian approach was
demonstrated in a previous paper \citep{paper1}.  In this paper we
have developed a pipeline for solving the inverse problem of inferring
the orbital parameters from a redshift curve.  We then compute the
redshift resolution needed to distinguish between redshift curves with
and without each relativistic perturbation included, thus simulating
the analysis of future observations.  The redshift resolution needed
to uncover each effect is a few times finer than the maximum
contribution of that effect.

In our treatment, we have assumed that frequency data is the only observable at hand. The prospects for detection would be improved if astrometric information were included in the fitting procedure. Information provided by astrometry is particularly potent in constraining the angular Keplerian elements, and would alleviate the burden placed on spectroscopy or pulse timing. 

A further way in which the detection prospects could be improved would be 
to obtain data from multiple orbits. 
The analysis of frequency data from stars or pulsars with periods $< 1$ yr would not only benefit from the steep period-dependence of the relativistic effects, but would also boost our chances of resolving cumulative effects. Precession effects are naturally of this type.

Figure~\ref{fig:at_30000} illustrates the redshift accuracies required
for S2.  With current instrumentation, capable of $\sim10\kms$ accuracies,
detecting time dilation appears comfortably feasible.  Detecting space
curvature appears feasible with a modest improvement in redshift
resolution.  On the other hand, the discovery of a pulsar on an orbit
comparable to S2 could push the effective redshift accuracy to $<1\ms$
and, in principle, bring frame dragging and even quadrupole effects
within reach, as shown in Figure \ref{fig:at_acc_3}.

This paper, however, assumes a source of negligible mass in pure Kerr
spacetime.  The finite mass of S2 (being $<10^{-5}\Mbh$) would perturb the
redshift by a similar factor, via its effect on the motion of the supermassive black hole. This would be much smaller than the
space-curvature effect, but more than the frame-dragging contribution. A potentially much more serious problem, however, is the Newtonian perturbations
due to other mass in the Galactic centre region.  This Newtonian
``foreground'' is not necessarily fatal -- the very specific
time-dependence of the relativistic effects may enable them to be
extracted from under a larger Newtonian perturbation 
\citep{nbody} -- but further
research is needed to assess this.

If sources inwards of S2 are discovered --- something we may hope for
from the E-ELT and the Square Kilometre Array (SKA) --- 
the prospects for relativity improve. 
\citet{PfahlLoeb2004} argue that there may be 100 pulsars with orbital 
periods less than 10 years, although the number may be much smaller
\citep[e.g.][]{2009arXiv0909.1318M}.
A large fraction of these are expected to be found by the SKA.
Not only do the relativistic effects get stronger as we move closer to the black hole, but the Newtonian
perturbations are likely to weaken.  Figures~\ref{fig:at_acc_10} and
\ref{fig:at_acc_1} shows the orbital periods at which one would need
to find stars or pulsars given redshift accuracies of $10\kms$ and
$1\kms$ respectively.  If stars with a period of less than one year
are discovered, the prospects become very exciting indeed.

\section*{Acknowlegements}
We thank Antoine Klein and Daniel D'Orazio for discussion and comments. 
DM was supported by grants AST-0807910 (NSF) and NNX07AH15G (NASA).
\pagebreak
\bibliographystyle{apj}
\bibliography{ms.bib}

\newpage

\begin{table}[h]
\begin{center}
\begin{tabular}{| l || l | l | l | }\hline
   Effect & Orbit      & Light path \\
    \hline
   Classical       & $P^{-1/3}$ &  \\
   Time dilation   & $P^{-2/3}$ &  \\
   Schwarzschild   & $P^{-1}$   & $P^{-1}$ \\
   Frame-dragging  & $P^{-3/4}$ & $P^{-5/3}$ \\
   Quadrupole      & $P^{-5/3}$ & $P^{-5/3}$ \\
  \hline 
\end{tabular}
\end{center}
\caption{Relativistic effects considered in this paper, and how
  they scale with the orbital period.}
\label{effect_grouping}
\end{table}

\begin{table}[!h]
\begin{center}
\begin{tabular}{| c || c || c | c | c | }\hline
   Effect & Required redshift accuracy & \multicolumn{3}{|c|}{Required orbital period (yr)}  \\
          & for S2 & \multicolumn{3}{|c|}{for given redshift accuracy}  \\
   & & 10$\kms$ & $1\kms$ & $30\cms$\\
   \hline
  Time dilation & $\sim 60\kms$ & $\sim 50$ & $\gg P_{S2}$ & $ \gg P_{S2}$\\
  \hline
  Schwarzschild & $\sim 3\kms$ & $\sim 13$ & $\sim 30$ & $ \gg P_{S2}$\\
  \hline
  Frame dragging & $\sim 10\ms$ & $\sim 0.5$ & $\sim 0.8$ & $ > P_{S2}$\\
  \hline
  Quadrupole & $\sim 50\cms$ & $\sim 0.04$ & $\sim 0.1$ & $\sim P_{S2}$\\
  \hline
\end{tabular}
\end{center}
\caption{Summary of the observational thresholds at which different
  relativistic effects are exposed, assuming a source of negligible
  mass in a pure Kerr spacetime.}
\label{results_summarised}
\end{table}

\newpage


\end{document}